\documentclass[fleqn]{article}

\begin{document}
\leftline{\Large\bf Universality and individuality in a neural code}
\bigskip\bigskip
\noindent {\large Elad Schneidman,$^{1,2}$ Naama Brenner,$^3$ Naftali 
Tishby,$^{1,3}$\\
Rob R. de Ruyter van Steveninck,$^3$ and William Bialek$^3$}
\bigskip

\leftline{$^1$Institute for Computer Science,}
\leftline{Center for Neural Computation and}
\leftline{$^2$Department of Neurobiology}
\leftline{Hebrew University}
\leftline{Jerusalem 91904, Israel}
\leftline{$^3$NEC Research Institute}
\leftline{4 Independence Way}
\leftline{Princeton, New Jersey 08540, USA}
\bigskip
\leftline{elads/tishby@cs.huji.ac.il}
\leftline{brenner/ruyter/bialek@research.nj.nec.com}
\bigskip

\bigskip\bigskip\hrule\bigskip\bigskip

\noindent The problem of neural coding is to understand
how sequences of action potentials (spikes) are related
to sensory stimuli, motor outputs, or (ultimately)
thoughts and intentions. One clear question is whether
the same coding rules are used by different neurons, or
by corresponding neurons in different individuals. We
present a quantitative formulation of this problem using
ideas from information theory, and apply this approach
to the analysis of experiments in the fly visual system.
We find significant individual differences in the
structure of the code, particularly in the way that
temporal patterns of spikes are used to convey
information beyond that available from variations in
spike rate. On the other hand, all the flies in our
ensemble exhibit a high coding efficiency, so that every
spike carries the same amount of information in all the
individuals. Thus the neural code has a quantifiable
mixture of individuality and universality. 

\vfill\newpage

\leftline{\large\bf Introduction}
\bigskip

When two people look at the same scene, do they see the
same things? This basic question in the theory of
knowledge seems to be beyond the scope of experimental
investigation. An accessible version of this question is
whether different observers of the same sense data have
the same neural representation of these data: how much
of the neural code is universal, and how much is
individual? To approach this problem we must give a
quantitative definition of similarity or distance among
neural codes. 

The problem of comparing neural codes has some analogies
to the problem of comparing among amino acid sequences
in proteins. In the neural code, sequences of action
potentials stand for sensory inputs or motor outputs; in
the genetic code sequences of nucleotides stand for
amino acids which in turn encode the three dimensional
structure of proteins. Just as motifs of several amino
acids collectively can encode structural elements in
proteins, patterns of several action potentials across
time or across a population of cells can have a special
meaning. For proteins, sequences can be similar because
they have a recent common ancestor; alternatively,
sequences can be equivalent functionally even if they
are distinguishable, since convergent evolution has led
to very different sequences that encode proteins with
similar structures and functions. For the neural code we
show that these notions of distinguishability and
functional equivalence can be quantified using ideas
from information theory \cite{shannon,pierce61}. In
particular this approach does not require a metric
either on the space of stimuli or on the space of neural
responses; all notions of similarity emerge from the
statistical structure of the neural responses. 

We apply these methods to analyze experiments on an
identified motion sensitive neuron in the fly's visual
system, the cell H1 \cite{hausen}. Many invertebrate
nervous systems have cells that can be named and
numbered \cite{ident}, and in many cases the total
number of neurons involved in representing a portion of
the sensory world is quite small, so that destruction of
individual neurons can have a substantial impact on
behavior (see, for example, Ref.
\cite{hausen-wehrhahn}). In these cases the neural
representation of sensory information is especially
accessible, precisely because it is localized to a small
set of identified cells. On the other hand, if a large
fraction of neurons is identifiable it might seem that
the question of whether different individuals share the
same neural representation of the visual world would
have a trivial answer. 

Far from trivial, we shall see that the neural code even
for identified neurons in flies has components which are
common among flies and significant components which are
individual to each fly. The existence of identified
neurons thus does not preclude the expression of
individuality in neural representations; we should
expect that all neural circuits, both vertebrate and
invertebrate, express a degree of universality and a
degree of individuality. For H1 we quantify these ideas,
and we hope that the methods we introduce will be
applicable more generally. In the interest of making the
discussion more accessible we have confined mathematical
arguments to the Methods section; in the main text we
make use of analogies to provide some intuition for what
the information theoretic quantities are measuring. 

\bigskip
\leftline{\large\bf Distinguishing among flies}
\bigskip

We place our discussion in the context of the
experiments shown in Fig. 1. Nine different flies are
shown precisely the same movie, which is repeated many
times for each fly; as we show the movie we record the
action potentials from the H1 neuron. The details of the
stimulus movie should not have a qualitative impact on
the results, provided that the movie that is
sufficiently long and rich to drive the system through a
reasonable and natural range of responses. Figure 1b 
makes clear that qualitative features of the neural 
response on long time scales ($\sim 100\,{\rm ms}$) are 
common to almost all the flies, and that some aspects of
the response are reproducible on a (few) millisecond
time scale across multiple presentations of the movie to
each fly. Nonetheless the responses are not identical in
the different flies, nor are they perfectly reproduced
from trial to trial in the same fly. The most obvious
difference among the flies is in the average spike rate,
which varies from 22 to 63 spikes/s among our ensemble
of flies. But beyond that, how should we quantify the
similarity or difference among the neural responses?

One way is to imagine that each spike train is a point
in an abstract space, and that there is a metric on this
space. Considerable effort has gone into the definition
of metrics that are plausible biologically and tractable
computationally \cite{waterman,vp97}, and these methods
have been used widely (for applications in neurobiology
(see \cite{vp97,vp96,MacLeod-etal-98}), but all
approaches based on metric spaces have several problems.
First, the metric is imposed by the investigator and
does not emerge from the data. Second, even within a
plausible class of metrics there are arbitrary
parameters, such as the relative distance cost of moving
vs. deleting a spike. Finally, it is not clear that our
intuitive notion of similarity among neural responses
(or amino acid sequences) is captured by the
mathematical concept of a metric. 

In contrast, information theory \cite{shannon} provides
a method for quantifying directly the differences among
the {\em sources} of the spike trains. Imagine that we
record multiple speakers reading from the same text, in
the same way that we record the activity of neurons from
different flies responding to the same sensory inputs.
There are many possible speakers, and we are shown a
small sample of the speech signal: how well can we
identify the speaker? If we can collect enough data to
characterize the distribution of speech sounds made by
each speaker then we can quantify, in bits, the average
amount of information that a segment of speech gives
about the identity of the speaker. Further we can
decompose this information into components carried by
different features of the sounds. Following this
analogy, we will measure the information that a segment
of the neural response provides about the identity of
the fly, and we will ask how this individuality is
distributed across different features of the spike
train. Apparently large differences in spike rate are
surprisingly uninformative, but the temporal patterns of
spikes allow for much more efficient discrimination
among individuals. The largest differences among
individuals are in the way that these patterns are
associated with specific stimulus features. 

We discretize the neural response into time bins of size
$\Delta t = 2\,{\rm ms}$, which is also the time
required to draw one frame of the stimulus movie. At
this resolution there are almost never two spikes in a
single bin, so we can think of the neural response as a
binary string, as in Fig. 1c-d. We examine the response
in blocks or windows of time having length $T$, so that
an individual neural response becomes a binary `word'
$W$ with $T/\Delta t$ 'letters'. Clearly, any fixed
choice of $T$ and $\Delta t$ is arbitrary, and so we
explore a range of these parameters. 

The distribution of words used by a particular fly's H1
in response to the stimulus movie, $P^{\rm i} (W)$ for
the ${\rm i^{th}}$ fly, tells us about the `vocabulary'
of that cell. Figure 1f shows that different flies
`speak' with similar but distinct vocabularies. From 
these distributions $P^{\rm i} (W)$ we can quantify the
average information that a single word of length $T$
gives about the identity of the fly, $I(W \rightarrow
{\rm identity};T)$ [see Eq. (\ref{infW}) in Methods].
Thus, we measure how well we can discriminate between
one individual and a mixture of all the other
individuals in the ensemble, or effectively how `far'
each individual is from the mean of her conspecifics. 

The finite size of our data set prevents us from
exploring arbitrarily long words, but happily we find
that information about identity is accumulating at more
or less constant rate $R$ well before the undersampling
limits of the experiment are reached (Fig. 2a). Thus
$I(W \rightarrow {\rm identity}; T) \approx R(W
\rightarrow {\rm identity}) \cdot T$; $ R(W \rightarrow
{\rm identity}) \approx 5 $ bits/s, with a very weak
dependence on the time resolution $\Delta t$. Since the
mean spike rate can be measured by counting the number
of 1s in each word $W$, this information includes the
differences in firing rate among the different flies.

Even if flies use very similar `vocabularies,' they may
differ substantially in the way that they associate
words with particular stimulus features. In our
experiments the stimulus runs continuously in a loop, so
that we can specify the stimulus precisely by giving the
time relative to the start of the loop; in this way we
don't need to make any assumptions about which features
of the stimulus are important for the neuron, nor do we
need a metric in the space of stimuli. We can therefore
consider the word $W$ that the ${\rm i^{th}}$ fly will
generate at time $t$. This word is drawn from the
distribution $P^{\rm i}(W|t)$ which we can sample, as in
Fig. 1c--e, by looking across multiple presentations of
the same stimulus movie. In parallel with the discussion
above, we can now ask for the average information that a
word $W$ provides about identity given that it was
observed at a particular time $t$. This depends on the
time $t$ because some moments in the stimulus are more
informative than others, as is obvious from Fig. 1. The
more natural quantity is an average over all times $t$,
which is the average information that we can gain about
the identity of the fly by observing a word $W$ at a
known time $t$ relative to the stimulus, $I(\{W,t\}
\rightarrow {\rm identity}; T)$ [see Eq.
(\ref{infWands}) in Methods]. 

Figure 2b shows a plot of 
$I(\{W,t\} \rightarrow {\rm identity}; T)/T$ 
as a function of the observation time 
window of size $T$. Observing both the spike train and 
the stimulus together provides $32\pm 1~{\rm bits/s}$ 
about the identity of the fly. This is more than six 
times as much information as we can gain by observing 
the spike train alone, and corresponds to gaining one 
bit in $\sim 30\,{\rm ms}$. Correspondingly, a typical 
pair of flies in our ensemble can be distinguished 
reliably in $\sim 30\,{\rm ms}$. This is the time scale 
on which flies actually use their estimates of visual 
motion to guide their flight during chasing behavior 
\cite{land-collett}, so that the neural codes of 
different individuals are distinguishable on the time 
scales relevant to behavior.

\bigskip
\leftline{\large\bf Spike rates and information rates}
\bigskip

Having seen that we can distinguish reliably among
individual flies using relatively short samples of the
neural response, it is natural to ask about the origins
and implications of these individual differences.
Perhaps the most obvious question is whether the
substantial differences in the code among the different
neurons have an impact on the ability of these cells to
convey information about the visual stimulus. As
discussed in Refs. \cite{science97,strong98}, the rate
at which the neural response provides information about
the visual stimulus, $R^{\rm i} (W \rightarrow s(t);
T)$, is determined by the same probability distributions
$P^{\rm i}(W|t)$ as before [see Eq's.
(\ref{stiminf1}--\ref{stiminf4}) in Methods]. Again we
note that our estimate of the information rate itself is
independent of any metric in the space of stimuli, nor
does it depend on assumptions about which stimulus
features are most important in the code. 

Figure 3a shows that the flies in our ensemble span a
range of information rates from $R^{\rm i}(W \rightarrow
s(t))\approx 50$ to $\approx 150\,{\rm bits/s}$. This
threefold range of information rates is correlated with
the range of spike rates, so that each of the cells
transmits nearly a constant amount of information per
spike, $2.39 \pm 0.24\,{\rm bits/spike}$. The error bar
in this case ($\pm 0.24\,{\rm bits/spike}$) is a
standard deviation across the ensemble of flies, not a
standard error of the mean: the number of bits per spike
transmitted by H1 (under these stimulus conditions) is
constant from fly to fly within $10\%$, despite three
fold variations in total spike rate.

Although information rates are correlated with spike
rates, this does not mean that information is carried by
a ``rate code'' alone. Rate coding usually is
distinguished from ``timing codes'' in which the
detailed temporal structure of the spike train plays a
crucial role. In particular, our computation of the
information carried by the spike train includes
automatically any contribution from temporal patterns,
but to demonstrate that these patterns are important we
must show that this information is more than we expect
just by summing the contributions of the individual
spikes that make up the patterns. This `single spike
information' can also be thought of as the information
conveyed by temporal modulations in the spike rate; see
Ref. \cite{brenner99} and Eq. (\ref{singlespikeinfo}) in
Methods. For all the flies in our ensemble, the total
rate at which the spike train carries information is
substantially larger than the `single spike'
information---$2.39$ vs. $1.64$ bits/spike, on average.
This extra information, as defined in Eq. (\ref{extra})
and illustrated in Fig. 3b, is carried in the temporal
patterns of spikes. 

The fact that the information per spike is constant
across the ensemble of flies means that cells with
higher spike rates are not generating extra spikes at
random, but rather each extra spike is equally
informative about the visual stimulus. The {\em
capacity} of the code to carry information is quantified
by the entropy rate ${\cal S}^{\rm i}_{\rm total}$ of
the distribution of neural responses [see Eq's.
(\ref{stiminf1},\ref{entrate}) in Methods], and is
different in different flies. It is natural
\cite{spikes,rieke93} to define the efficiency of the
code as the fraction of this capacity which is used to
convey information about the visual stimulus,
$\epsilon^{\rm i} = R^{\rm i}(W \rightarrow s(t))/{\cal
S}^{\rm i}_{\rm total}$. Like the information per spike,
this efficiency is nearly constant across the ensemble
of flies, $\epsilon = 0.59 \pm 0.05$, at $\Delta t =
2\,{\rm ms}$, with a very weak dependence on $\Delta t$
\cite{strong98}.

\bigskip
\leftline{\large\bf A universal codebook?}
\bigskip

Even though flies differ in the structures of their
neural responses, distinguishable responses could be
functionally equivalent, as with distinct amino acid
sequences that fold to the same protein structure. 
It might therefore be that all flies could be endowed
(genetically?) with a universal or consensus codebook
that allows each individual to make sense of her own
spike trains, despite the differences from her
conspecifics. Thus we want to ask how much information
we lose if the identity of the flies is hidden from us,
or equivalently how much each fly can gain by knowing
its own individual code. 

The codebook for any individual fly can be thought of as
a probabilistic mapping from neural responses or words
back into the space of visual stimuli \cite{88}. The
information conveyed by the spike train quantifies the
specificity of this mapping: the `tighter' the
distribution of stimuli consistent with a given response
the more information is conveyed. If the neural codes
used by different flies are different, then these
conditional distributions in stimulus space are also
different. If we don't know the identity of the fly, all
we can do is to associate each neural response with a
distribution of stimuli that corresponds to an average
over the individuals, and this distribution necessarily
is broader than any of the individual distributions. As
a result, we have less information about the visual
stimulus, as summarized by Eq. (\ref{codingloss}) in the
Methods. 

Intuitively, the greater the differences among the
neural responses of different flies, the more visual
information we will lose if we don't know the identity
of the individual. On the other hand, these differences
mean that the neural response provides information about
individual identity, so that information gained about
identity is information lost about the stimulus if we
use a universal codebook. This intuitive connection is
made precise by Eq. (\ref{codingloss+identinf}) in the
Methods. As a practical matter, this means that the
answer to our question about the efficacy of a universal
decoder is contained in the results of Fig. 2. The
result is that, on average, not knowing the identity of
the fly limits us to extracting only $64$ bits/s of
information about the visual stimulus. This should be
compared with the average information rate of $92.3$
bits/s in our ensemble of flies: knowing her own
identity allows the average fly to extract $44 \%$ more
information from H1. Further analysis shows that each
individual fly gains approximately the same relative
amount of information from knowing its personal
codebook. 

\vfill\newpage
\bigskip
\leftline{\large\bf The nature of the `personal' bits}
\bigskip

Thus far we have analyzed the differences among the
neural codes of different flies, and how much extra
information a fly can extract by knowing it's individual
codebook. It is natural to ask what is being ``said'' by
these extra bits, characterizing more explicitly the
mapping from neural responses back to stimulus space for
the different flies. 

For each neural response $W$ we can look back through
the entire experiment and accumulate the motion
trajectories that lead up to the response, and these
provide samples from the distribution of stimuli
conditional on the response as described above. Because
the space of trajectories has many dimensions, this
distribution is difficult to visualize, and so we focus
here on the means of these distributions. This is a
generalization of the reverse correlation or spike
triggered average method \cite{spikes}: rather than
looking at the average stimulus that leads to a single
spike, we look at the average stimulus that leads to the
responses $W$, which can consist of a pattern of spikes
and empty intervals \cite{88}.

In Fig. 4 we show the average waveforms of the stimulus
velocity preceding a specific binary word in the spike
trains of flies 1 and 6. As fly 6 spike trains convey
almost 3 times more information about the stimulus, one
might have speculated that the same word was used in
completely different stimulus contexts for the two
flies. In fact, the differences are in the details and
not in the general picture: spikes stand for pulses of
positive velocity (as in Fig. 4b), long silent intervals
stand for negative velocities (as in Figs. 4a\&c), and
the largest differences among the flies are in the
widths, latencies and amplitudes of the pulses;
combinations of spikes and intervals then lead to very
different trajectories (as in Fig. 4d). For the fly
which conveys less information, spikes are associated
with larger positive velocities (Fig. 4b) and silences
are associated with (slightly) larger negative
velocities (Fig. 4a); thus, these elementary responses
come closer to exhausting the dynamic range of the
inputs. Conversely, the more informative spike train
covers the dynamic range of inputs with a greater
variety of composite responses.

\bigskip
\leftline{\large\bf Discussion}
\bigskip

One obvious difference between invertebrate and
vertebrate nervous systems is the existence of
identified neurons in invertebrates. The identifiability
of invertebrate neurons sometimes has been interpreted
to mean that these smaller nervous systems are hard
wired automata; indeed the optomotor system of flies has
been held up as a clear example of this extreme view. In
this view, individuality plays no role, and it should
even be possible to average the results of experiments
on corresponding neurons in different individuals. For
vertebrates, substantial individuality arises through
development and learning, and there are few if any
identified neurons; at best vertebrates have
identifiable modules consisting of hundreds or thousands
of neurons, such as the columns in visual cortex.
Against this clear dichotomy it is worth remembering
that even genetically identical single celled organisms
exhibit individuality in their sensory--motor behavior
\cite{spudich-koshland}. 

In the present work we have tried to the quantify the
individuality of the neural code used by a single neuron
in the fly visual system. On the one hand, this
individuality is sufficient to allow discrimination
among individuals on time scales of relevance to
behavior. Correspondingly, each individual fly would
lose a significant amount ($\sim 30\%$) of the visual
information carried by this neuron if it `knew' only the
codebook appropriate to the whole ensemble of flies. On
the other hand, these differences among the codebooks of
different flies seem to be matters of detail. Although
different flies extract very different amounts of
information from the same visual inputs, all the flies
achieve a high and constant efficiency in their encoding
of this information. From previous work it is known that
the visual system of an individual fly exhibits
substantial changes in coding strategy as it adapts to
different ensembles of inputs. Rather than converging on
the same information rates in different flies, these
adaptation processes seem to converge on codes of
uniformly high efficiency, supporting the idea that
efficiency of representation is a `design principle' for
the system \cite{barlow61}. 

On average the flies in our ensemble have neural codes
in which a substantial amount of information is carried
by patterns of spikes. This antiredundancy or synergy
among spikes \cite{brenner99} is reduced substantially
if we are forced to use a universal codebook.
Mathematically this loss of synergy in the universal
codebook is related to the fact that the rate at which
we gain information about the identity of the fly (Fig.
2b) increases with window size out to $T_{\rm c}\sim 10
\,{\rm ms}$: discrimination among flies is enhanced by
being able to see patterns of spikes in windows of size
$T_{\rm c}$, implying that the way these patterns are
used to encode visual information is unique to each
individual. Each individual fly thus gains nearly $50\%$
more information through the use of a code in which
patterns of spikes carry extra information, and more
than half of this is lost if the fly does not have
knowledge of its own identity. Not only is spike timing
important for the neural code, but the way in which
timing is used is specific to each individual. 

\bigskip\bigskip\hrule\bigskip
\bigskip
\leftline{\large\bf Methods}
\bigskip

\parindent =0pt

{\bf Flies, neural recording and stimulus generation}

Recordings were made from the H1 neuron using standard
methods: the fly was immobilized in wax, a tungsten
microelectrode was inserted through a small hole at the
back of the fly's head, and H1 was identified through
its response properties; spikes were detected with a
window discriminator. The stimulus was a rigidly moving
pattern of vertical bars, randomly dark or bright, with
average intensity ${\bar I}\approx 100 {\rm mW/(m^2
\cdot sr)}$, displayed on a Tektronix 608 high
brightness display; bar widths were set equal to the
horizontal lattice spacing (interommatidial angle) of
the compound eye. The fly viewed the display through a
round diaphragm, showing approximately 30 bars. Frames
of the stimulus pattern were refreshed every 2 ms, and
with each new frame the pattern was displayed at a new
position. This resulted in an apparent horizontal motion
of the bar pattern, which is suitable to excite the H1
neuron. The pattern position was defined by a
pseudorandom sequence, simulating a diffusive motion or
random walk. We draw attention to three points relevant
for the present analysis: (1) The flies are freshly
caught female {\em Calliphora,} so that our `ensemble of
flies' approaches a natural ensemble and is not
restricted to a highly inbred laboratory stock. (2) In
each fly we identify the H1 cell as the unique spiking
neuron in the lobula plate that has a combination of
wide field sensitivity, inward directional selectivity
for horizontal motion, and contralateral projection. (3)
Recordings are rejected only if raw electrode signals
are excessively noisy or unstable; in particular we do
not select for flies that exhibit mean spike rates
(spontaneous or driven) in a predefined range. \medskip 

{\bf Definition of $I(W \rightarrow {\rm identity}; T)$}

Imagine that we record the response of the H1 neuron
from one fly, but we don't know which one. A priori
there are $N$ equally likely possibilities. Once we
observe the spike train for some time $T$ our
uncertainty is reduced, and hence we gain information
about the identity of the fly from which we are
recording. The average information that an individual
word provides about the fly's identity is
\begin{equation} I(W \rightarrow {\rm identity}; T) =
\sum_{{\rm i}=1}^N P_{\rm i} \sum_W P^{\rm i} (W) \log_2
\left[ {{P_{\rm i}P^{\rm i}(W)}\over {P^{\rm ens}(W)}}
\right] ~{\rm bits,} \label{infW} \end{equation} where
$P_{\rm i}=1/N$ is the a priori probability that we are
recording from fly ${\rm i}$ and $P^{\rm ens}(W)$ is the
probability that any fly in the whole ensemble of flies
would generate the word $W$, \begin{equation} P^{\rm
ens}(W) = \sum_{{\rm i}=1}^N P_{\rm i} P^{\rm i} (W) .
\label{Pens} \end{equation} The measure $I(W \rightarrow
{\rm identity}; T)$ has been discussed by Lin
\cite{Lin-91} as the `Jensen--Shannon divergence'
$D_{\rm JS}$ among the distributions $P^{\rm i} (W)$. We
recall that the problem of finding a measure of
similarity among distributions is not simple; obvious
choices such as the Kullback--Leibler divergence are not
symmetric, and may have spurious technical requirements
such as absolute continuity of one distribution with
respect to the others. Lin proposed $D_{\rm JS}$ as a
way of getting around these difficulties, and he showed
that $D_{\rm JS}$ can be used to bound other measures of
similarity, such as the optimal or Bayesian probability
of identifying correctly the origin of a sample (as in
forced choice psychophysical discrimination
experiments). Here $D_{\rm JS}$ arises not just as an
interesting possible measure of similarity (see
also \cite{El-Yaniv-FiTi-97}), but as the unique answer
to the question of how much information a sample
provides about its source. \medskip \vfill\newpage 

{\bf Definition of $I(\{W,t\} \rightarrow {\rm identity}; T)$} 

By analogy with Eq. (\ref{Pens}) we define the
distribution of words used at time $t$ by the whole
ensemble of flies. 

\begin{equation} P^{\rm ens}(W|t) = 
\sum_{{\rm i}=1}^N P_{\rm i} P^{\rm i} (W|t), 
\end{equation} 

and by analogy with Eq. (\ref{infW}) we
can measure the information that the word $W$ observed
at known $t$ gives us about the identity of the fly,

\begin{equation} I(W \rightarrow {\rm identity} | t; T)=
\sum_{{\rm i}=1}^N P_{\rm i} \sum_W P^{\rm i} (W|t)
\log_2 \left[ {{P_{\rm i}P^{\rm i}(W|t)}\over 
{P^{\rm ens}(W|t)}} \right] ~. 
\end{equation} 

The more natural quantity is an average over all times
$t$, 

\begin{equation} I(\{W,s(t)\} \rightarrow {\rm identity}; T) = 
\langle I(W \rightarrow {\rm identity} |t; T) \rangle_{t} 
~{\rm bits,} 
\label{infWands}
\end{equation} 

where $\langle \cdots \rangle_t$ denotes an average over
$t$. \medskip 

{\bf Definition of $I^{\rm i} (W \rightarrow s(t); T)$}

The discussion here follows Refs.
\cite{science97,strong98}. The entropy of the
distribution of words, 

\begin{equation} S_{\rm total}^{\rm i} (T) = 
-\sum_W P^{\rm i} (W) \log_2 P^{\rm i}(W) \,{\rm bits}, 
\label{stiminf1} 
\end{equation}

measures the capacity of the neuron to transmit
information. At each time $t$ we can define the entropy
of the conditional distribution $P^{\rm i}(W|t)$, which
measures the noise in the response to repeated
presentations of the same movie, 

\begin{equation} S_{\rm noise}^{\rm i} (t; T) = 
-\sum_W P^{\rm i}(W|t) \log_2 P^{\rm i}(W|t) \,{\rm bits.} 
\label{stiminf2} 
\end{equation} 

The information is the difference between the total
entropy of the cell's vocabulary and the average noise
entropy, 

\begin{equation} I^{\rm i} (W \rightarrow s(t); T) = 
S_{\rm total}^{\rm i} (T) - \langle S_{\rm noise}^{\rm 
i} (t; T) \rangle_t . 
\label{stiminf3} 
\end{equation}
 
For sufficiently large window of time $T$, we expect 
that to gain information in proportion to the duration 
of our observations, 

\begin{equation} I^{\rm i} (W \rightarrow s(t); T) 
\approx R^{\rm i}(W \rightarrow s(t)) \cdot T , 
\label{stiminf4} 
\end{equation} 

so that there is a well defined information rate 
$R^{\rm i}(W \rightarrow s(t))$. This asymptotic behavior is
observed in the data for values of $T$ that are relevant
to fly behavior. Similar behavior is observed for the
total entropy, 

\begin{equation} S_{\rm total}^{\rm i} (T) \approx 
{\cal S}_{\rm total}^{\rm i} \cdot T, 
\label{entrate} 
\end{equation} 

leading to the definition of coding efficiency discussed
in the text \cite{spikes,rieke93}. \medskip 

{\bf Information from single spikes} 

The discussion here follows Ref. \cite{brenner99}. In
the experiments the stimulus runs continuously for a
time $T_{\rm loop}$ and then repeats. When we observe a
single spike at time $t$ we learn something about the
stimulus in the neighborhood of this time, and if we
average this information over all possible arrival times
we obtain the average information carried by a single
spike. Because information is mutual, we can relate the
information that the spike provides about the stimulus
to the information that the stimulus provides about the
occurrence of a spike, but this is contained in the time
dependent firing rate or post--stimulus time histogram
for cell $\rm i$, $r_{\rm i}(t)$. After some algebra
\cite{brenner99}, the single spike information takes the
form of an integral that depends only on $r_{\rm i}(t)$, 

\begin{equation}
I^{\rm i}_{\rm one\ spike} = {1\over {T_{\rm loop}}}\int_0^{T_{\rm loop}}
dt \,{{r_{\rm i}(t)}\over{\bar r_{\rm i}}} \log_2 \left[
{{r_{\rm i}(t)}\over{\bar r_{\rm i}}} \right] \, {\rm bits,}
\label{singlespikeinfo}
\end{equation}

where $\bar r_{\rm i}$ is the average spike rate in cell
$\rm i$. If spikes were to carry information
independently, then each cell would transmit 
$R^{\rm i}_{\rm ind} = \bar r_{\rm i}I^{\rm i}_{\rm one\
spike}$ bits per second. If the total information rate 
$R^{\rm i}(W \rightarrow s(t))$ is smaller than this 
then spikes are redundant (on average) while if the 
total information rate is larger then there is synergy
\cite{brenner99} among the spikes and the extra
information must be carried in the temporal patterns of
spikes. We can quantify this extra information as a
fraction, 

\begin{equation}
F^{\rm i}_{\rm extra} = [R^{\rm i}(W \rightarrow
s(t)) - R^{\rm i}_{\rm ind}]/R^{\rm i}_{\rm ind} ,
\label{extra}
\end{equation}

as illustrated in Fig. 3b.
\medskip

{\bf Information loss with universal decoding}

If we observe the response of a neuron but don't know
the identity of the individual generating this response,
then we are observing responses drawn from the ensemble
distributions defined above, $P^{\rm ens}(W|t)$ and
$P^{\rm ens}(W)$. The information that words provide
about the visual stimulus then is 

\begin{equation}
I^{\rm mix}(W \rightarrow s(t); T) = \Bigg\langle \sum_W P^{\rm ens}
(W|t) \log_2 \left[ {{P^{\rm ens}(W|t)}\over {P^{\rm ens}(W)}}
\right] \Bigg\rangle_t \,{\rm bits.}
\end{equation}

On the other hand, if we know the identity of the fly to
be $\rm i$, we gain the information $I^{\rm i} (W
\rightarrow s(t); T) $ from above [Eq's.
(\ref{stiminf1}--\ref{stiminf3})]. The average
information loss is then 

\begin{equation}
I_{\rm loss}^{\rm avg} (W \rightarrow s(t); T)
= \sum_{{\rm i}=1}^N P_{\rm i} I^{\rm i} (W \rightarrow s(t); T)
- I^{\rm mix}(W \rightarrow s(t); T) .
\label{codingloss}
\end{equation}

After some algebra it can be shown that this average
information loss is related to the information that the
neural responses give about the identity of the
individuals, as defined above: 

\begin{eqnarray}
I_{\rm loss}^{\rm avg} (W \rightarrow s(t); T)
&=& I(\{W,t\} \rightarrow {\rm identity}; T)
 \nonumber\\
 &&\,\,\,\,\,\,\,\,\,\,\,\,\,\,
- I(W \rightarrow {\rm identity}; T) .
\label{codingloss+identinf}
\end{eqnarray}

\newpage 

{\bf Figure 1}. Different flies' spike trains and
word statistics. {\bf(a)} All flies view the same random
vertical bar pattern moving across their visual field
with a time dependent velocity, part of which is shown
(see Methods section for details). In the experiment, a
40 sec waveform is presented repeatedly, 90 times. {\bf
(b)} A set of 45 response traces to the part of the
stimulus shown in (a) from each of the 9 flies. The
traces are taken from the segment of the experiment
where the transient responses have decayed. Spike trains
from flies 1 and 6 are colored by red and blue,
respectively, which we will use as a color code for the
other parts of the figure.{\bf(c)} Example of
construction of the local word distributions. Zooming in
on a segment of the repeated responses of fly 1 to the
visual stimuli (see green rectangle in (b)), the fly's
spike trains are divided into contiguous 2 ms bins, and
the spikes in each of the bins are counted. E.g., we get
the 6 letter words that the fly used at time 3306 ms
into the input trace. {\bf(d)} Similar as (c) for fly 6.
{\bf(e)} The distributions of words that flies 1 and 6
used at time $t=3306\, {\rm ms}$ from the beginning of
the stimulus. The time dependent distributions,
$P^{1}(W|t=3306\, {\rm ms})$ and $P^{6}(W|t=3306\, {\rm
ms})$ are presented as a function of the binary value of
the actual 'word', e.g., binary word value $'17'$ stands
for the word $'010001'$. {\bf(f)} Collecting the words
that each of the flies used through all of the visual
stimulus presentations, we get the total word
distributions for flies 1 and 6, $P^1(W)$ and $P^6(W)$.

\bigskip

{\bf Figure 2}. Distinguishing one fly from others based
on spike trains. {\bf (a)} The average rate of 
information gained about the identity of a fly, given 
the distribution of words that it used throughout the
stimulus presentations, as a function of the word size
used. The information rate is saturated even before we
reach the maximal word length used; for more discussion
of word lengths see Methods. Following Figure 1, Red
marks are the average rate of information that the word
distribution of fly 1 give about its identity, compared
with the word distribution mixture of all of the flies.
The connecting line is used for clarification only. Blue
marks the results for fly 6, and the black marks the
average over all 9 flies. See methods for discussion 
of error bars calculation. {\bf (b)} Similar to the
computation done for (a), we can compute the average
amount of information that is gained about the identity
of the fly, give its word distribution at a specific
time, compared with the mixture of the word distribution
of all of the 9 flies. Averaging over all times, we get
the average amount of information gained about the
identity of fly 1 based on its time dependent word
distributions (red), fly 6 (blue), and the average over
the 9 flies (black).

\bigskip

{\bf Figure 3}. The information about the stimulus that 
a fly's spike train carries is correlated with firing
rate, and yet a significant part is in the temporal
structure. {\bf(a)} The rate at the H1 spike train
provides information about the visual stimulus is shown
as a function of the average spike rate, with each fly
providing a single data point (Fly 1 is marked by a red
point and Fly 6 by a blue one). The linear fit of the
data points for the 9 flies corresponds to a universal
rate of $2.39\pm 0.24\,{\rm bits/spike}$, as noted in
the text. {\bf(b)} The extra amount of information that
the temporal structure of the spike train of each of the
flies carry about the stimulus, as a function of the
average firing rate of the fly [see Eq. (\ref{extra}) in
Methods]. The average amount of additional information
that is carried by the temporal structure of the spike
trains, over the population is $45 \pm 17 \%$.

\bigskip

{\bf Figure 4}. What different flies mean by the same 
words. The word-triggered averages are shown for flies 1
(red) and 6 (blue) for 4 different words. Similar to the
computation of spike triggered averages, we compute the
average velocity profile of the movie presented to the
flies, preceding 7-letter binary words. {\bf(a)} The
average stimulus waveform preceding the word $'0000000'$
for Flies 1 (red) and 6 (blue), is shown as a function
of the time relative to the end of the word (shown in
actual time order on the right top side of the panel).
{\bf(b-d)} Word triggered averages for 3 other words,
reflecting that the waveforms haves similar rough
structure, and that the difference between the flies is
in the details. 

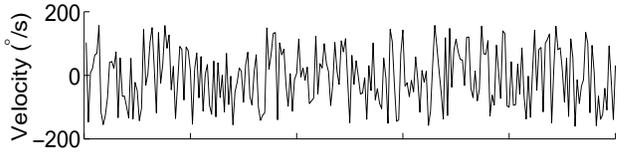
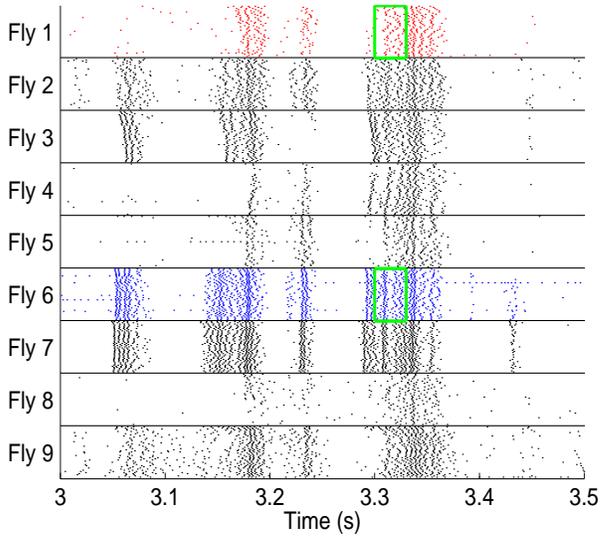
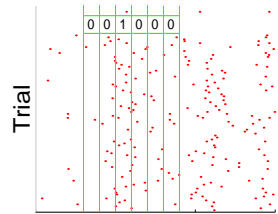
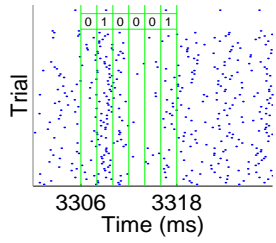
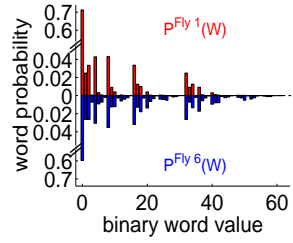
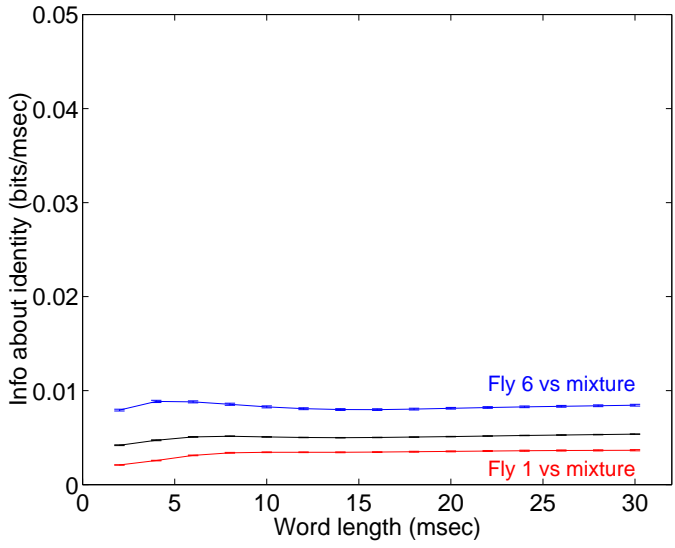
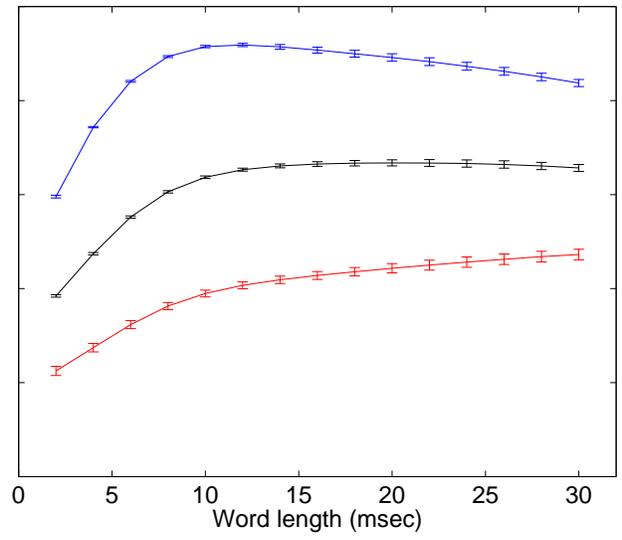
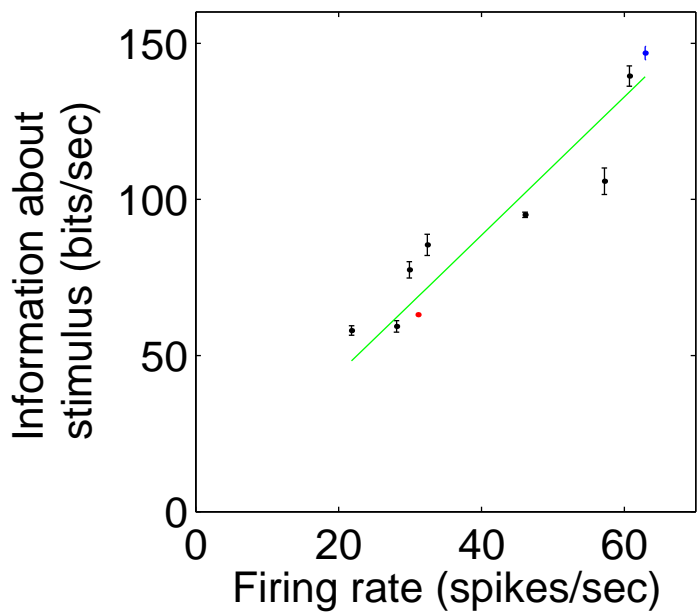
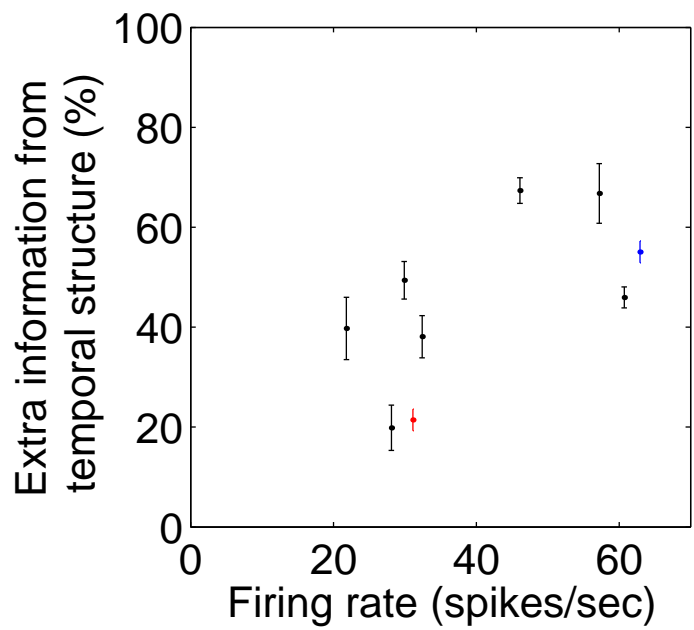
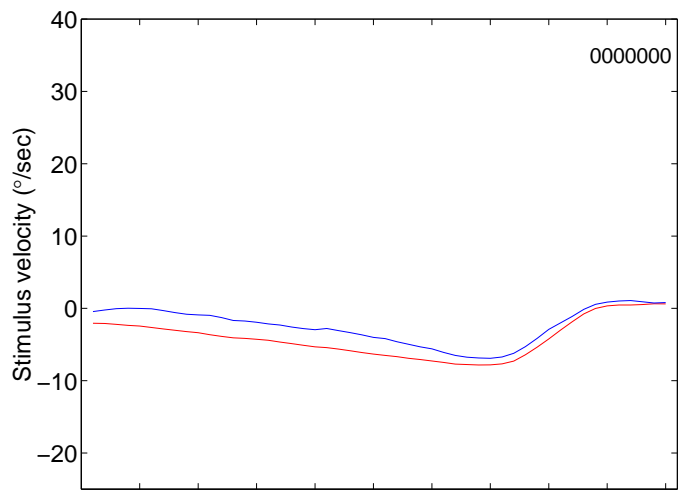
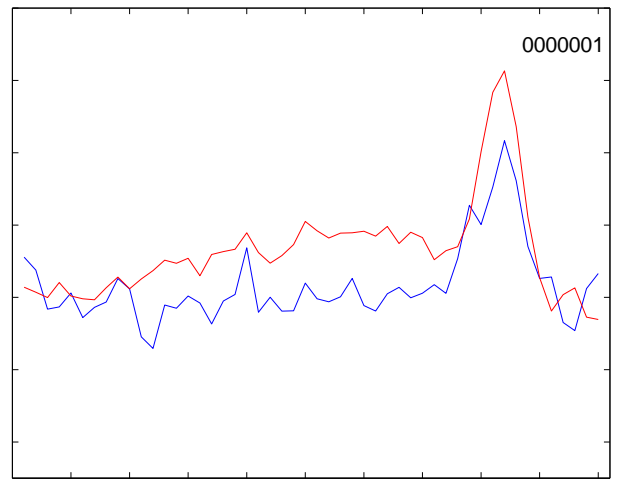
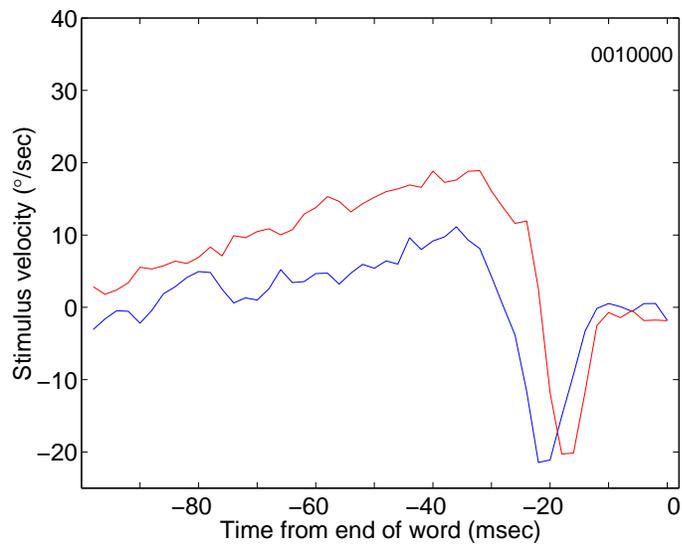
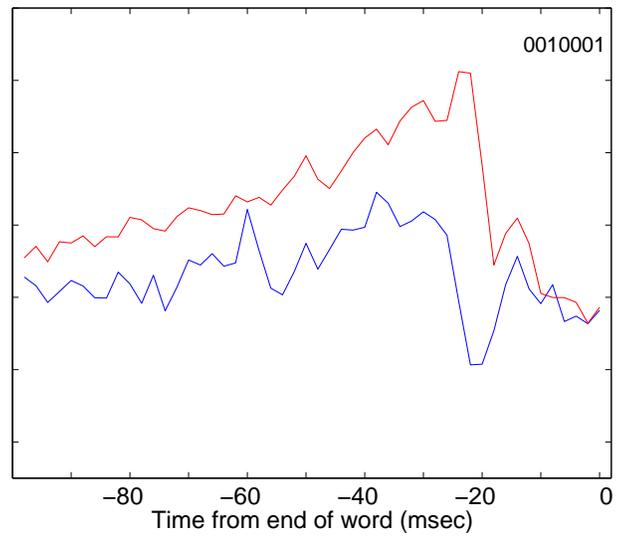
\end{document}